\documentclass[pre,twocolumn,amsmath,pacs]{revtex4}
\usepackage{graphicx}
\begin{document}

\title{
Epitaxial growth with pulsed deposition:\\
Submonolayer scaling and Villain instability
}
\author{Berit Hinnemann$^{1,2}$, Haye Hinrichsen$^3$, 
      and Dietrich E. Wolf$^1$}
\affiliation{$^1$ Institut f\"ur Physik, 
     Gerhard-Mercator-Universit{\"a}t Duisburg,
     47048 Duisburg, Germany}
\affiliation{$^2$ CAMP, Dept. of Physics, Technical University of Denmark,
        DK-2800 Lyngby, Denmark}
\affiliation{$^3$ Theoretische Physik, Fachbereich 8,
     Universit{\"a}t GH Wuppertal,
     42097 Wuppertal, Germany}
\date{April 17, 2002}


\begin{abstract}
  It has been observed experimentally that under certain conditions
  pulsed laser deposition (PLD) produces smoother surfaces than
  ordinary molecular beam epitaxy (MBE). So far the mechanism leading
  to the improved quality of surfaces in PLD is not yet fully understood. In
  the present work we investigate the physical properties of a simple
  model for PLD in which the transient mobility of adatoms and
  diffusion along edges is neglected. Analyzing the crossover from MBE
  to PLD, the scaling properties of the time-dependent nucleation
  density as well as the influence of Ehrlich-Schwoebel barriers we
  find that there is indeed a range of parameters where the surface
  quality in PLD is better than in MBE. However, since the improvement is weak 
  and occurs only in a small range of parameters we conclude that
  deposition in pulses alone cannot explain the experimentally
  observed smoothness of PLD-grown surfaces.
\end{abstract}

%

\maketitle
\parskip 1mm


\section{Introduction}

Pulsed Laser Depostion (PLD) is an increasingly used growth method
which is particularly suited for the fabrication of multilayer thin
films~\cite{ChriseyHubler}. The basic mechanism of this technique is
that the target material is ablated by a pulsed laser and then
deposited in pulses on a substrate so that many atoms arrive
at the surface simultaneously. Experimentally each pulse lasts for
about a few nanoseconds, and the time between two pulses is of the
order of seconds. PLD differs significantly from ordinary molecular
beam epitaxy (MBE), where atoms are deposited continuously.
Growth experiments show that the surfaces in both cases
have different morphologies. In particular it was observed that in
some cases PLD leads to smoother surfaces than MBE~\cite{Jenniches96}.
However, as there are only few theoretical studies so far, the
mechanism leading to smoother surfaces in PLD is not yet understood so
that it is impossible to predict the growth quality for various
materials and growth parameters. Therefore the aim of the present work
is to improve our understanding of PLD by investigating the growth
morphology for different growth parameters in a simple model.

Without going into experimental details it should be mentioned that
the physical conditions for PLD are far less well defined than for
MBE. The particles deposited may be atoms, clusters or even droplets.
They may arrive with energies ranging from $0.1$ eV to 1000 eV. If
very energetic particles arrive at the surface their kinetic energy
is converted into heat at the surface, changing locally the
effective mobility of the particles for a short time. In the present
work this transient mobility is neglected, i.e., it is assumed that
the particles arrive at the substrate surface with energies of about
$0.1$ eV which are of the same order of magnitude as in the case of
thermal deposition.

Recently Narhe {\em et al.} investigated the island statistics in the
coalescence regime for a PLD process where tin droplets are deposited
on a sapphire substrate~\cite{Narhe}. It was observed that the scaling
differs significantly from MBE due to the large fraction of multiple droplet
coalescence under pulsed vapor delivery. While these results are valid
for high deposition energies, the present work investigates a
different physical regime, namely pulsed deposition of individual
atoms at low energies and the formation of islands in two dimensions.

The model investigated in the present work is defined as a
solid-on-solid growth process on a square lattice of $L \times L$
sites with integer heights representing the configuration of the
adsorbed layer.  The particles are deposited in pulses with an
intensity $I$ which is defined as the number of particles per unit
area deposited per pulse. The duration of a pulse is assumed to be
zero and the transient enhancement of the mobility of freshly
deposited adatoms is neglected. The model is controlled by three
parameters, namely, the intensity $I$ of the pulses, the diffusion
constant $D$, and the average flux density of incoming particles $F$.
One of these parameters can be fixed by choosing the time scale so
that we can use $I$ and $D/F$ as independent parameters. The dynamic
rules are defined as follows. (i) In each pulse $IL^2$ atoms are
instantaneously deposited at random positions on the surface. (ii)
Between two pulses a time interval $\Delta t = I/F$ elapses, in which
adatoms diffuse on the surface with rate $D$. If Ehrlich-Schwoebel
barriers are present the rate at which particles hop down the
edge of an island is reduced. (iii) If two atoms at the same height
occupy neighboring sites they stick together irreversibly, forming the
nucleus of a new island or attaching to an already
existing island. In our model the influence of edge diffusion is
neglected so that islands grow in a fractal manner before they
coalesce. We present results from
kinetic Monte Carlo simulations on a
lattice of $400 \times 400$ sites.
Finite size effects are discussed in the Appendix.

The paper is organized as follows. In the following section we first
recall the properties of PLD without Ehrlich-Schwoebel barriers and
its crossover to MBE in the limit of very low intensities 
\cite{JensenNiemeyer,ConferencePaper}. In
Sec.~\ref{SecTimeNucleations} we investigate the scaling behavior of
the time-dependent nucleation density in PLD, extending our previous
analysis in Ref.~\cite{OurPRL}. In Sec.~\ref{SecSchwoebel} we study
the influence of Ehrlich-Schwoebel barriers on the morphology of
surfaces grown by PLD, finding a parameter range where layerwise
growth is improved compared to MBE.

\section{From MBE to PLD:\newline \hspace*{7mm} Crossover and Scaling} 
\label{SecMBEPLD}
%
%
\begin{figure}
\includegraphics[width=80mm]{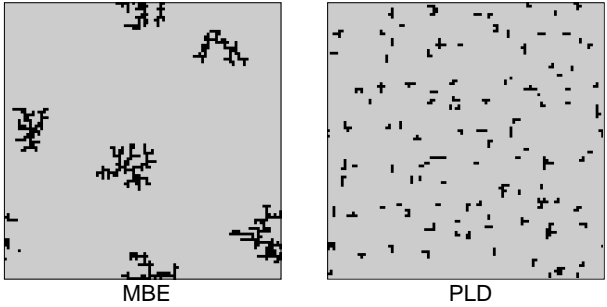}
\vspace{-3mm}
\caption{Molecular beam epitaxy (left) compared to pulsed laser
deposition (right) for $D/F=10^8$ and $I=0.01$. The figure shows
typical configurations after deposition of $0.05$ ML.}
\label{FigMBEPLD}
\end{figure}
If the intensity $I$ is very small PLD behaves essentially in the same
way as MBE. In fact, at the lowest possible intensity, where only one
atom per pulse is deposited, MBE and PLD are equivalent up to minor
statistical differences resulting from finite size effects (see
Appendix). Using lattices of $400^2$ sites we therefore restrict our
analysis to the range $D/F\leq10^7$ where these finite size effects
are negligible.

At high intensities the growth morphology of PLD differs significantly
from MBE. As shown in Fig.~\ref{FigMBEPLD}, there are much more
nucleations at an early stage, although the effective flux of incoming
particles is the same in both cases. The two regimes are separated by
a crossover at a certain intensity $I_c$, where the number
of deposited atoms per pulse is of the same order of magnitude as the
average adatom density in the corresponding MBE process. Obviously, if
the pulse intensity is much higher than the MBE adatom density, the
adatoms nucleate much faster forming many small islands. For this
reason PLD is expected to yield more homogeneous surfaces, which
explains the technological interest in this method.

Let us now study the crossover from MBE to PLD 
in more detail~\cite{JensenNiemeyer,ConferencePaper}. The adatom density in
MBE averaged over space and time is known to scale as $(D/F)^{2\gamma
-1}$~\cite{VPW92}, where

\begin{equation}
\label{gamma}
\gamma=\frac{1}{d_f+4} =\left\{
\begin{array}{ll}
1/6 & \text{for compact islands}\\
0.18 & \text{for DLA fractals \cite{Meakin88}}
\end{array}  \right. \,.
\end{equation}
Thus the crossover takes place at the critical intensity
\begin{equation}
I_c \propto (D/F)^{2\gamma -1}.
\label{CriticalIntensity}
\end{equation}
A quantity which distinguishes the two different growth modes shown in
Fig.~\ref{FigMBEPLD} is the average island distance. Performing
numerical simulations we realized that the scaling regime of this
quantity is not only restricted by finite-size effects but also by
lattice effects, which become relevant for high intensities as well as for low
values of $D/F$, where the average distance of islands is of
the order of a few lattice constants. Combining these bounds we find
that for a system with $400^2$ sites finite-size and lattice effects
are negligible in the parameter range $10^4\leq D/F \leq 10^7$ and
$I\leq 10^{-2}$.

\begin{figure}
\vspace{-10mm}
\includegraphics[width=75mm, angle=270]{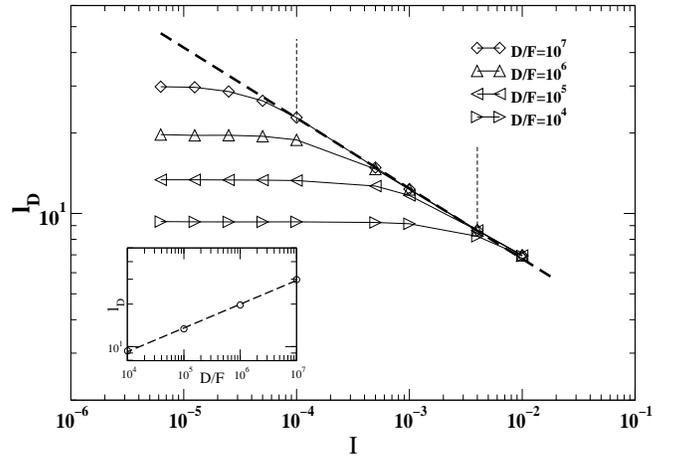}
\vspace{-5mm}
\caption{Average island distance $\ell_D$ versus pulse intensity $I$
  for different values of $D/F$. The data points between the two
  dotted vertical lines are used to determine the exponent $\nu$. The
  dashed line is the corresponding power-law fit with the slope
  $-\nu=-0.26(1)$. The inset shows the saturation levels of $\ell_D$ for
  small $I$ as a function of $D/F$. The dashed line has slope $\gamma
  = 0.17(1)$.} 
\label{FigScaling}
\end{figure}

Fig.~\ref{FigScaling} shows the average island distance as a function
of the intensity for various values of $D/F$ at a fixed
coverage of $0.2$ monolayers (ML). For low intensities
$I<I_c$ the island distance depends only on $D/F$. Plotting these
saturation values versus $D/F$ (shown in the inset of the figure)
one recovers the well-known power law for the island distance in MBE
\begin{equation}
  \ell_D\propto (D/F)^{\gamma}
\label{IslandDistanceMBE}
\end{equation}
with $\gamma =0.17(1)$. This estimate lies between the values for 
compact growth and diffusion-limited aggregation (see Eq. (\ref{gamma})), 
supporting the assumption that the islands are characterized by 
an effective fractal dimension $1.6<d_f<2$. 

\begin{figure}
\vspace{-11mm}
\includegraphics[width=72mm, angle=270]{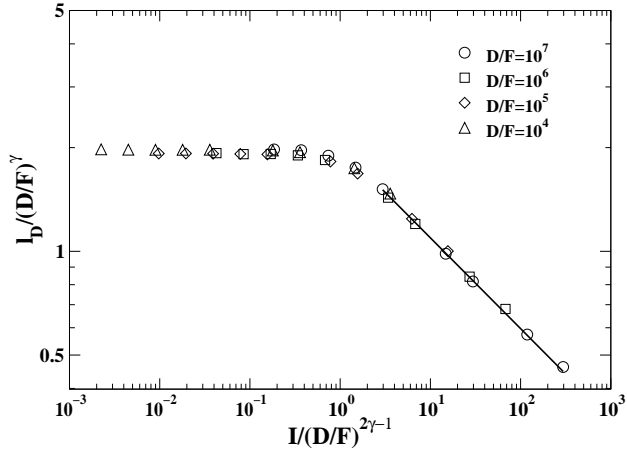}
\vspace{-5mm}
\caption{Data collapse of the curves in Fig.~\ref{FigScaling} according to
  (\ref{PowerScaling}) using the exponents $\gamma=0.17$ and
  $\nu=0.26$. The knee of the curves marks the crossover from MBE- to
  PLD-like behavior.} 
\label{FigScalingCollapse}
\end{figure}

For intensities $I>I_c$ the island distance is independent of $D/F$ 
and can be described by a power law 
\begin{equation}
  \ell_D \propto I^{-\nu}
\label{Nu}
\end{equation}
with an exponent $\nu = 0.26(1)$. The independence of the island 
distance of $D/F$ is a result of the high density of adatoms which 
nucleate so quickly that they do not make use of their full mobility given by $D/F$.

Combining Eqs.~(\ref{IslandDistanceMBE}) and~(\ref{Nu}) 
we obtain the scaling form~\cite{ConferencePaper}
\begin{equation}
\ell_D\propto (D/F)^{\gamma}\cdot h(I/I_c) \,,
\label{PowerScaling}
\end{equation}
where $h$ is a scaling function with the asymptotic behavior
\begin{equation}
\label{PowerScalingFunction}
h(y)\sim \left\{
\begin{array}{ll}
                \mbox{const.} & \mbox{for } y\ll 1\\
                y^{-\nu}  & \mbox{for } y\gg 1.
           \end{array} \right.
\end{equation}
As shown in Fig.~\ref{FigScalingCollapse}, this scaling form 
leads to a convincing data collapse. Moreover, the independence 
of $D/F$ at high intensities together with Eq.~(\ref{CriticalIntensity}) implies that
\begin{equation}
\nu =\frac{\gamma}{1-2\gamma} \simeq \left\{
\begin{array}{ll}
0.25 & \text{for compact islands}\\
0.28 & \text{for DLA fractals}
\end{array}  \right. \,.
\label{PowerNu}
\end{equation}
confirming our numerical estimate $\nu=0.26(1)$.

\section{Time-dependent nucleation density}
\label{SecTimeNucleations}

\begin{figure}
\vspace{-2mm}
\includegraphics[width=60mm, angle=270]{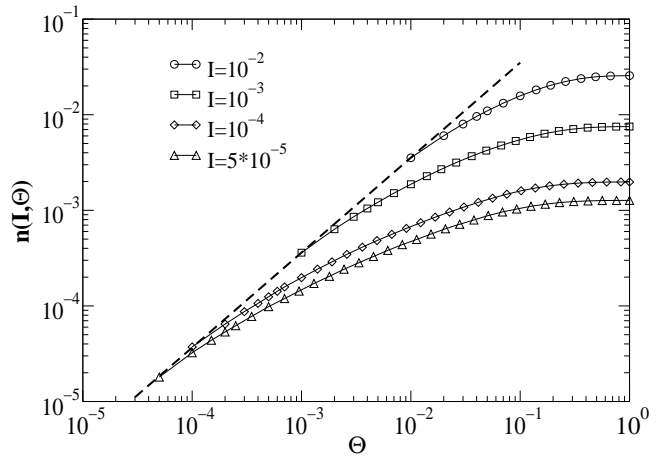}
\caption{The nucleation density at the bottom layer 
versus monolayer time during the deposition 
of one monolayer. The dashed line has the slope~1.}
\label{RawResults}
\end{figure}

\begin{figure}
\includegraphics[width=60mm, angle=270]{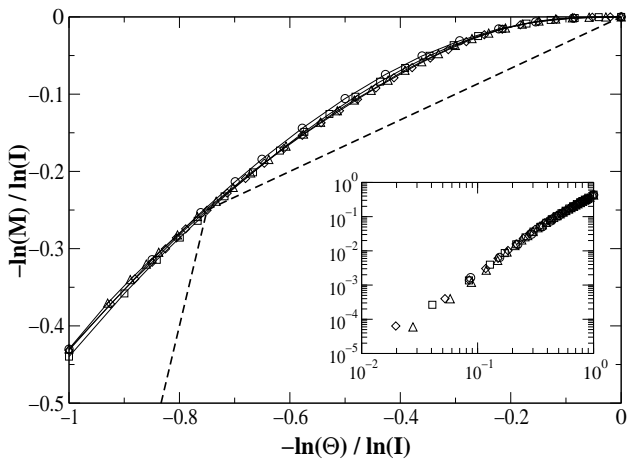}
\caption{Data collapse according to the scaling form (\ref{ScalingResult}). 
The inset shows a double-logarithmic plot of $\ln M/\ln I$ vs.
$\ln \Theta/\ln I$. The line is slightly 
curved with local slopes varying from 2.4 to 2.0.}
\label{FinalScaling}
\end{figure}

In a recent paper~\cite{OurPRL} we investigated the time-dependent 
nucleation density in PLD, reporting an unusual type of scaling
behavior. In order to avoid the crossover to MBE, we considered the 
limit $D/F \to \infty$ where $I_c=0$. In this limit all adatoms 
nucleate or attach to existing islands before the next pulse 
arrives. Fig.~\ref{RawResults} shows a log-log plot of the 
the nucleation density at the bottom layer $n(I,\Theta)$ 
as a function of the coverage $\Theta=Ft$ for various 
intensities $I$. Obviously this quantity does not display 
ordinary power law scaling since it is impossible to collapse 
the curves by shifting them horizontally and vertically. However, 
in Ref.~\cite{OurPRL} we observed that the normalized nucleation density
\begin{equation}
\label{Mdef}
M(I,\Theta) = n(I,\Theta)/n(I,1)
\end{equation}
obeys an unusual {\em logarithmic} scaling law of the form
\begin{equation}
\ln M(I,\Theta) \;\simeq\;
\left(\ln I\right) \,g(\ln \Theta/\ln I).
\label{ScalingResult}
\end{equation}
This scaling form was also proposed by Kadanoff {\it et al.} 
and Tang in the context of multiscaling 
in self-organized criticality~\cite{Kadanoff88,Tang89}. 
As shown in Fig.~\ref{FinalScaling} this scaling form leads to a
convincing data collapse. More recently the same type of 
scaling has also been observed in one-dimensional systems, 
so that we can rule out logarithmic corrections at the 
marginal dimension of random walks as the origin for 
this type of non-conventional scaling~\cite{Lee,Krapivsky}.

In Fig.~\ref{FinalScaling} the left terminal points of the curves at
$g(1)=0.44(2)$ can be used to determine $\gamma$ as follows. It is
known that the nucleation density after the first pulse grows linearly
with the intensity~\cite{OurPRL} while the nucleation density after
completion of one monolayer grows as $n(I,1)\propto I^{-2 \nu}$ (see
Eq.~(\ref{Nu})). Therefore, the normalized nucleation
density after the first pulse scales as $M(I,I) \propto I^{1-2\nu}$ so
that 
\begin{equation}
g(1) = 1-2\nu = \frac{1-4\gamma}{1-2\gamma} \,.
\end{equation}
Solving this equation we obtain $\gamma=0.179(4)$
in agreement with the estimate in Fig.~\ref{FigScaling}.

In order to understand how this unusual scaling behavior for PLD
crosses over to the ordinary power-law scaling of MBE let us now turn
to the case of finite $D/F$, where the system is characterized by a
typical length scale $\ell_0 \sim (D/F)^{1/4}$. Generalizing the
results of Ref.~\cite{OurPRL} we consider islands with an arbitrary
fractal dimension, i.e., we regard $\gamma$ as a free parameter.  

In MBE the nucleation density is known to exhibit ordinary power law
scaling of the form 
\begin{equation}
n(\ell_0,\Theta) = \ell_0^{-2} \, f(\Theta \ell_0^2) \,,
\end{equation}
where $f$ is a scaling function with the asymptotic behavior~\cite{Tang93}
\begin{equation}
\label{MBE_AsymptoticScaling}
f(z) \propto \left\{
\begin{array}{ll}
z^3 & \text{ for } 0 \leq z \ll 1 \\
z^{1/3} & \text{ for } 1 \ll z \lesssim z_{\rm max}\,.
\end{array}
\right. 
\end{equation}
The upper bound $z_{\rm max}$ is determined by the condition that the
whole surface is covered by islands so that no further nucleation in
the respective layer is possible \cite{VPW92}. Because of $I_c\sim
(D/F)^{2\gamma-1}$ the length scale $\ell_0$ is related to the
crossover intensity by 
\begin{equation}
\ell_0 \sim I_c^{-1/(4-8\gamma)}\,.
\end{equation}
Using this scaling form we obtain the expression
\begin{equation}
M(I_c,\Theta) = \left\{
\begin{array}{ll}
\Theta^3 I_c^{-\frac{3-1/3}{2-4\gamma}} & \text{ for } 0 \leq z \ll 1 \\
\Theta^{1/3} & \text{ for } 1 \ll z \lesssim z_{\rm max} \,,
\end{array}
\right. 
\end{equation}
where $z=\Theta I_c^{-1/(2-4\gamma)}$. Taking the logarithm and
extrapolating to $z=\ell_0^2$ we arrive at 
\begin{equation}
\label{PuzzleGamma}
\frac{\ln M(I_c,\Theta)}{\ln I_c}=
\left\{ 
\begin{array}{ll}
 3\frac{\ln\Theta}{\ln I_c}+\frac{4}{6\gamma -3} &  \mbox{ for }\frac{1}{2-4\gamma}
\ll \frac{\ln\Theta}{\ln I_c} \\[2mm]
  \frac{\ln\Theta}{3\ln I_c} & \mbox{ for }0\ll\frac{\ln\Theta}{\ln I_c}
\ll\frac{1}{2-4\gamma}\,.
\end{array}
 \right. 
\end{equation}
In the limit $I \simeq I_c  \rightarrow 0$ (i.e., $D/F\to \infty)$ 
the crossover between both regimes becomes sharper and converges 
to a piecewise linear curve, which is shown in Fig.~\ref{FinalScaling} 
as a dashed line. The crossover point is located at
\begin{eqnarray}
\ln\Theta_c/\ln I_c &=& 1/(2-4\gamma), \nonumber\\
\ln M(I_c,\Theta)/\ln I_c &=&1/(6-12\gamma)\,.
\label{crossover}
\end{eqnarray}
Surprisingly, the crossover point lies on the collapsed curves for PLD
within numerical errors. This is plausible for the following reasons.
On the one hand the PLD-curve must be an upper bound for the MBE-curve
since the island density in PLD is always larger. On the other hand,
if the gap between the two extrapolated curves did not close at the
crossover between PLD- and MBE-behavior, it would imply that there is
an additional characteristic length in the system, for which we have
no evidence.

Furthermore, we note that the scaling function itself 
roughly follows a power law
\begin{equation}
\label{parabola}
g(z) \simeq g(1) \, z^{\beta}\,.
\end{equation}
As shown in the double-logarithmic inset of Fig.~\ref{FinalScaling}, 
the  effective exponent $\beta$ varies between 2.4 and 2.0. 
In order to verify this estimate we derive the exponent $\beta$ 
from Eq.~(\ref{crossover}) by assuming that the crossover point 
lies exactly on the collapsed curves for PLD. This leads to 
an expression for $\beta$ in terms of the exponent $\gamma$, namely
\begin{equation}
\beta = \frac{\ln (6-24\gamma)}{\ln (2-4\gamma)}
 \simeq \left\{
\begin{array}{ll}
2.4 & \text{ for compact islands}\\
2.15 & \text{ for DLA fractals }
\end{array}  \right. \,.
\label{BetaResult}
\end{equation}
Since this is just the range in which the numerical values for $\beta$
vary, we are led to the conclusion that the effective fractal
dimensions of the islands for low and high coverages are different.
For low coverages the islands are spaced relatively far apart so that
the growth is DLA-like while for high coverages the islands coalesce
and become more and more compact.

\section{Influence of Ehrlich-Schwoebel barriers}
\label{SecSchwoebel}

In most experimental situations interlayer transport is
reduced by an additional energy barrier $E_{\rm ES}$ which the atoms have to
overcome when hopping down an edge of an island~\cite{Schwoebel}. This
barrier is called {\em Ehrlich-Schwoebel} (ES) barrier and is of the
order of about $0.1$ eV for metals~\cite{SmilauerSchwoebel}. A useful
measure for the influence of this barrier is the {\em Schwoebel
  length} $\ell_{\rm ES}$, which is defined as~\cite{KrugSchwoebel} 
\begin{equation}
  \ell_{\rm ES} = a\cdot\mbox{exp}\left[\frac{E_{\rm ES}}{k_{\rm
      B}T}\right]\,,
\label{Schwoebellength}
\end{equation}
where $a=1$ denotes the lattice constant. For MBE it is known 
that Ehrlich-Schwoebel barriers impede interlayer transport. 
This leads to a growth instability which was predicted and 
first theoretically investigated by Villain~\cite{Villain91} 
followed by others~\cite{Siegert} and has also been observed 
experimentally~\cite{SchwoebelObservations}.  In contrast to MBE, not
much is known about the influence of Ehrlich-Schwoebel barriers on
PLD. However, it has been observed experimentally that PLD leads to
better growth results than MBE in certain situations where
Ehrlich-Schwoebel barriers are present~\cite{Jenniches96}.

The influence of ES barriers on PLD is twofold. On the one hand, for
high intensities many adatoms are deposited on the same island, which
should increase the influence of ES barriers. On the other hand, in
PLD the islands are much smaller so that adatoms tend to leave an
island very quickly, thereby reducing the influence of ES barriers.
Therefore, the question arises whether the experimentally observed
improved quality of layer-by-layer growth in PLD can be related to a
reduced influence of ES barriers and whether it is possible to choose
the parameters of the model in such a way that PLD produces 
smoother surfaces than MBE.

This question has been addressed previously by Schinzer et
al. \cite{Biehl}. They studied the special case where one intense 
pulse (0.23 ML) is deposited at the beginning of each monolayer, while
the remaining atoms are deposited with a continuous flux. The aim was
to check an idea of Rosenfeld {\it et al.}~\cite{Rosenfeld}, that it should
improve the smoothness of the surface, if the islands are forced to be
much smaller than the diffusion length of the atoms deposited after
the pulse. Surprisingly, Schinzer {\it et al.}~\cite{Biehl} found that this
is not the case in their simulation. On the contrary, in the presence
of the pulse the surface became rougher.

A key concept for describing the influence of ES-barriers is 
the time it takes before the first
nucleation in the second layer takes place. Layer-by-layer growth
requires that nucleations in the second layer do not start
significantly before the first layer is completed~\cite{Villain91}.
In the case of MBE the study of the second-layer nucleation time
turned out to be very useful in order to predict the growth mode for a
given set of parameters~\cite{Tersoff94,Krug00}. In the following we
demonstrate that this concept can be successfully applied to PLD as
well. To this end we first investigate the growth behavior of PLD in
the limit of $D/F \rightarrow\infty$. It will be shown that for any
finite Ehrlich-Schwoebel barrier MBE produces a better layer-by-layer
growth than PLD in agreement with the finding of Schinzer et
al.\cite{Biehl}. Then we compare MBE and PLD for different 
values of $I$ and $D/F$ over a wide range of Ehrlich-Schwoebel barriers. It
turns out that there is indeed a regime where PLD produces better
growth results than MBE.


\subsection{Infinite $D/F$}
\label{SubPnuc}

\begin{figure}
\vspace*{-8mm}
\includegraphics[width=90mm]{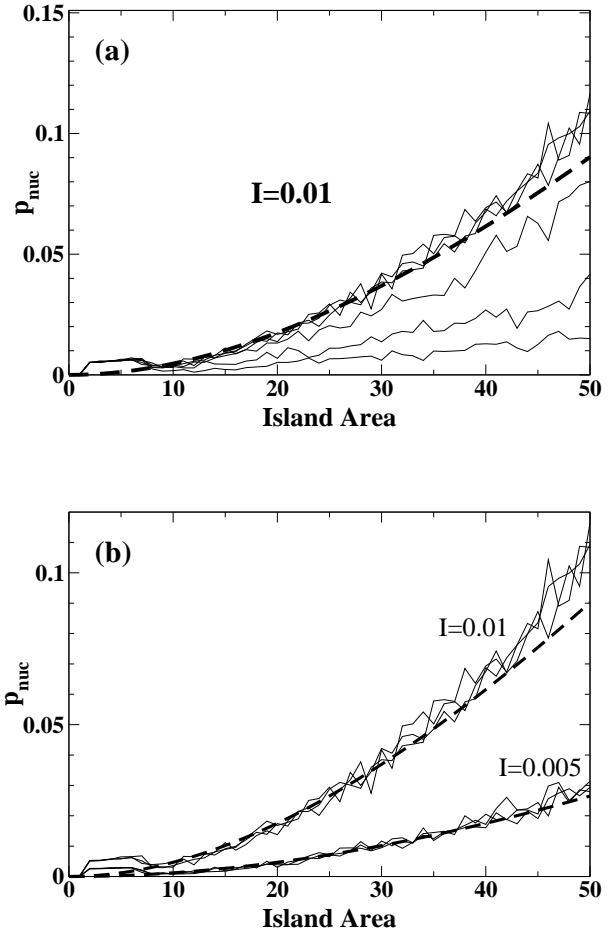}
\caption{
Numerical measurement of the second-layer nucleation probability 
compared to the theoretical prediction. After each pulse 
$p_{\rm nuc}$ is estimated as the fraction of islands with 
area $A$, where a second-layer nucleation happened for the 
first time, averaging over several runs. The island area 
is restricted to $A\leq 50$, where coalescence does not yet 
play a role. (a) The second layer nucleation probability 
$p_{\rm nuc}$  for $I=0.01$ versus island size $A$ for 
different Ehrlich-Schwoebel barriers in the limit $D/F \to \infty$. 
From bottom to top the Schwoebel lengths are  
$\ell_{ES}=10^0,10^1,10^2,10^3,10^4,10^5$. 
(b)  Plot of $p_{\rm nuc}$ versus~$A$ for two different 
intensities and for high Ehrlich-Schwoebel barriers 
$\ell_{\rm ES}=10^3, 10^4, 10^5$.}
\label{FigPnuc}
\end{figure}

We start by explaining why in the limit $D/F \rightarrow \infty$ PLD
does not delay the Villain instability in spite of the island size
reduction compared to MBE, provided that the ES-barrier is sufficiently high
and not leaky e.g. at kink sites. In this limit the time 
scales are separated as follows. Because of the high ES 
barrier the nucleation time of adatoms on top of islands 
is much smaller than their residence time, while for 
$D/F \rightarrow \infty$ the residence time is in turn 
much smaller than the time interval between two pulses. 
This means that we can restrict our analysis to a single pulse.

\begin{figure}
\vspace*{-8mm}
\includegraphics[width=90mm]{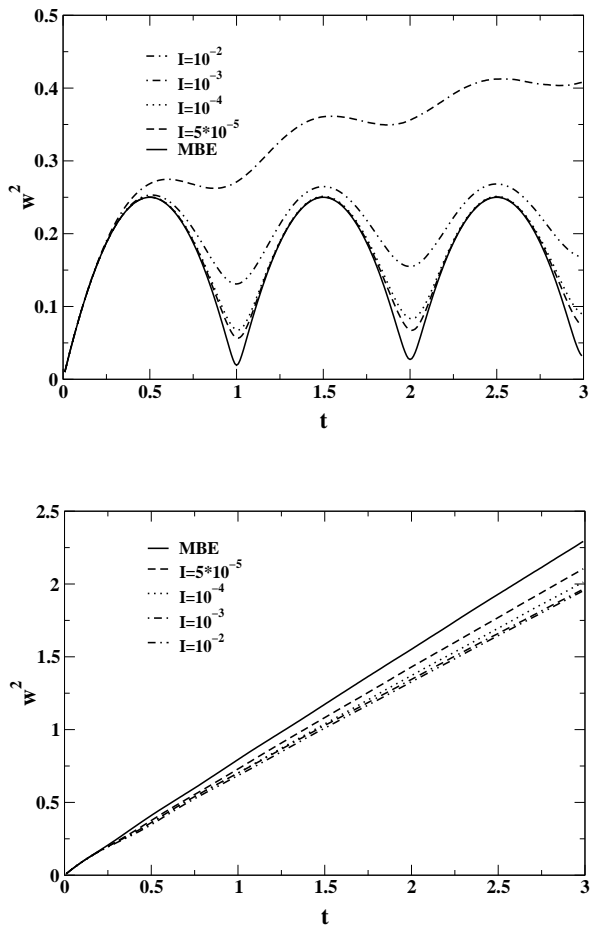}
\caption[Squared surface width for MBE and PLD]{(a) Squared surface 
width versus time for MBE and PLD for different intensities. 
The growth parameters are $D/F=10^8$ and $\ell_{\rm ES}=1$. 
(b) Squared surface width versus intensity for $D/F=10^8$ 
and $\ell_{\rm ES}=10^4$.}
\label{fig:comp_examples}
\end{figure}

The second layer nucleation probability $p_{\rm nuc}(I,A)$ is defined as the
probability that there is at least one nucleation event on islands 
with area $A$ after deposition of a pulse with intensity $I$. For 
high ES barrier this means that during a single pulse at least two 
atoms have to be deposited on the same island. Obviously, the 
probability of depositing $k$ atoms on an island with area $A$ 
during one pulse is given by a Poisson distribution with average $I\cdot A$,
\begin{equation}
p_k= \frac{\left(I\cdot A\right)^k}{k!}e^{-I \cdot A}.
\label{eq:poisson}
\end{equation}
Hence the nucleation probability can be expressed as
\begin{eqnarray}
\nonumber
p_{\rm nuc}(I,A) &=& 1-p_0-p_1 = 1-e^{-IA} \, \left(1+IA\right) \\
&\simeq& I^2A^2 + O(I^3A^3) \,.
\label{eq:pnuc}
\end{eqnarray}
Fig.~\ref{FigPnuc}a shows the numerical results for $p_{\rm nuc}$ for 
$I=0.01$ and different ES-barriers. As can be seen, for increasing 
ES barrier the measured curves approach the predicted one, 
(\ref{eq:pnuc}), which is shown as a dashed line. The agreement 
is good for $\ell_{ES}\geq 10^3$, where the nucleation time is 
much shorter than the residence time. However, the agreement 
is not convincing for island areas $A<10$, where the discrete 
lattice spacing starts to play a role. Moreover, the upper curves 
deviate for $A>40$ due to coalescence of large islands. 
This explanation is supported by the 
results shown in Fig.~\ref{FigPnuc}b, where two different intensities
are compared. Indeed, for the lower intensity $I=0.005$, where
coalescence starts at larger island sizes, the measured and the
predicted curves agree much better. Thus we conclude that the second
layer nucleation probability $p_{\rm nuc}$ is adequately described by
(\ref{eq:pnuc}).

With Eq.~(\ref{eq:pnuc}) we can now answer which of the opposing
trends in PLD -- island size reduction and increasing nucleation
probability on top of islands -- will dominate. In the limit
$D/F\rightarrow\infty$, where the critical intensity $I_c$ tends to
zero, the island area scales as $A\sim \ell_D^2 \sim I^{-2\nu}$  for
all intensities. Together with Eq.~(\ref{eq:pnuc}) one obtains to
leading order  
\begin{equation}
p_{\rm nuc}\propto I^{2(1-2\nu)}.
\end{equation}
Using the previous estimate $\nu \simeq 0.26$
the exponent is given by $2(1-2\nu)\simeq 1$.  Therefore, the second
layer nucleation probability $p_{\rm nuc}$ grows with increasing intensity,
enhancing the Villain instability. Contrarily, MBE with infinite 
$D/F$ always shows perfect layer-by-layer growth, even for 
high but finite ES-barriers. Thus, we conclude that in the 
limit $D/F\rightarrow\infty$ PLD cannot improve layer-by-layer growth. In
order to apply this result to practical situations, one has to find
the lower boundary for $D/F$, below which the behavior is different.
This question will be addressed now.


\subsection{Comparison of PLD and MBE for finite $D/F$}
\label{SubComparison}
\begin{figure*} 
\hspace*{-5mm}
\includegraphics[width=130mm]{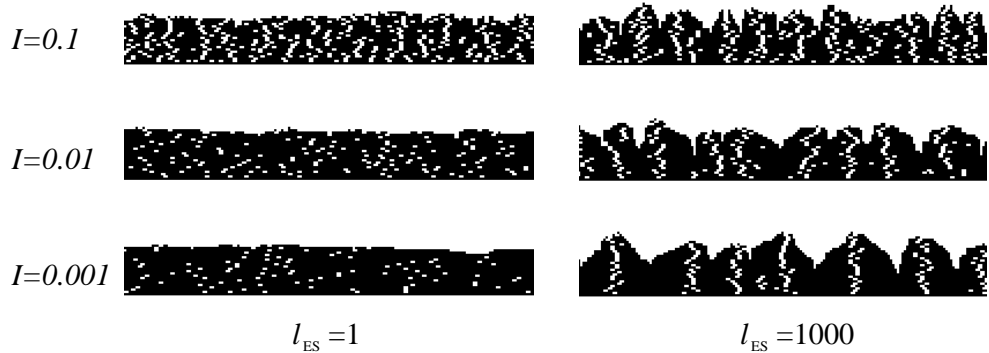}
\caption{1+1-dimensional PLD after deposition of $20$ ML for
different pulse intensities. The nucleation sites are shown as white
dots. Left: Without Ehrlich-Schwoebel barrier the roughness
increases with increasing intensity. Right: If the Ehrlich-Schwoebel
barrier is high enough, the autocorrelation among nucleation sites
is more pronounced than for $\ell_{\rm ES} = 1$, and
the roughness decreases with increasing intensity.}  
\label{FigNucDemo}
\end{figure*}

For perfect layer-by-layer growth the squared surface width 
$w^2=({\langle  h\rangle}^2-\langle h^2 \rangle)$ is known 
to oscillate between zero for completed monolayers and $1/4$ 
for half monolayers~\cite{Wolf95}. Without ES barriers such 
oscillations can be seen in MBE as well as in PLD. As shown 
in Fig.~\ref{fig:comp_examples}a, they are most pronounced 
in MBE while in PLD they become more and more damped as the 
intensity increases. Moreover, it can be seen that in PLD 
without ES barriers the roughness always increases with 
increasing pulse intensity.

Fig.~\ref{fig:comp_examples}b shows the corresponding result 
for a very high Ehrlich-Schwoebel length $\ell_{\rm ES}=10^4$. 
There are no oscillations since the surface roughens very quickly 
due to the Villain instability~\cite{Villain91}. Surprisingly, 
the width is now maximal for MBE and {\em decreases} with increasing 
pulse intensity, i.e., the trend is reversed. This reversal can 
even be observed visually by monitoring the interface at different 
intensities, as shown in Fig~\ref{FigNucDemo}.

Let us now investigate this reversal in more detail. Since the 
oscillations in PLD are extremely weak or not present at all, 
a numerical measurement of the damping time (such as in 
Ref.~\cite{Kallabis97} for MBE) is not feasible. Instead we 
monitored the squared surface width after deposition of a 
two monolayers -- a coverage which is also relevant for 
experimental applications~\cite{Jenniches96}. Although the 
choice of two monolayers is arbitrary and does not permit a 
rigorous quantitative analysis, this criterion is very simple 
and tells us for which parameters PLD produces smoother surfaces than MBE.

\begin{figure}
\vspace*{-7mm}
\includegraphics[width=70mm, angle=270]{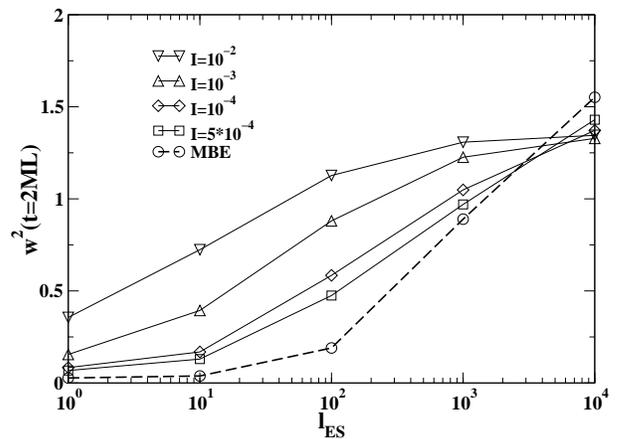}
\caption{The squared surface width at $\Theta=2$ML versus the 
Schwoebel length for MBE and PLD with various intensities 
and for $D/F=10^8$. The curve for MBE is dashed.}
\label{fig:PLDMBE}
\end{figure}

The roughness after deposition of two monolayers is plotted in
Fig.~\ref{fig:PLDMBE}. In agreement with previous results for
MBE~\cite{Brendel}, all curves increase monotonously, i.e., the
roughness increases with increasing Ehrlich-Schwoebel barrier. However, the
curves for MBE and PLD cross each other. The typical ES barrier, where
this crossing takes place, varies roughly as $\ell_{\rm ES} \approx
(D/F)^{1/2}$. This observation is in agreement with our previous
result  that for $D/F \to \infty$ surfaces grown by MBE are always
smoother compared to PLD. 

Thus there is indeed a range of parameters where PLD with a 
high intensity produces smoother surfaces than MBE, even if the atoms are
deposited with thermal energy. However, the Ehrlich-Schwoebel 
barriers are unphysically large (typical experimental values 
are $\ell_{ES}\approx 1\ldots 10$). Therefore we believe that 
nonthermal energy deposition effects are important for explaining 
the experimentally observed growth improvement in PLD.

Nevertheless it is interesting to address the question, what causes the
crossing of the curves in Fig.~\ref{fig:PLDMBE}. As we have explained, second
layer nucleation becomes more likely the larger the pulse intensity is
in spite of the decreasing island size. Therefore one would always
expect the roughness to increase with pulse intensity, as it is indeed the
case for sufficiently small Ehrlich-Schwoebel barriers. For 
$\ell_{\rm ES}\to\infty$, however, there is no interlayer 
transport and the roughness is the same as in random deposition, 
irrespective of the pulse intensity. Thus, the curves in 
Fig.~\ref{fig:PLDMBE} are expected to saturate eventually at 
the same roughness. Thus the observed reversal results from 
an interplay of a large but finite ES barrier and a finite value of $D/F$.

To understand the reversal let us consider how the Villain instability
unfolds for large but finite ES barriers. Depending on the intensity,
the first pulse leads to the formation of many small islands. After
the first pulse there are two temporal regimes of different roughening
behavior. Because of the Villain instability effective uphill currents
lead to a quick formation of large mounds. As new nucleations
preferentially take place on top of these mounds, the nucleations are
vertically aligned~\cite{Somfai}, as shown on the r.h.s. of
Fig.~\ref{SubComparison}. Therefore, the number and the lateral size
of the mounds is essentially determined by the initial configuration
of small islands after the first pulse. This determines at least for
transient times also the height of the mounds and therefore the
roughness of the surface. In our simulation model we do not observe
slope selection. In a more realistic case one would expect that the mounds
grow until their edges reach a critical slope where the uphill current
becomes zero. Then the process enters a different temporal
regime where the mounds compete with one another, leading to an effective
coarsening process where the roughness increases only slowly.  

By increasing the pulse intensity second layer nucleations become more
likely so that the Villain instability is accelerated. However, the
typical roughness from where on the process enters the second temporal
regime of slow coarsening depends mainly on the initial density of
islands after the first pulse. Therefore, this typical roughness
{\em decreases} with increasing intensity. After sufficiently long
time this effect dominates and leads to the observed crossings of the
curves in Fig.~\ref{fig:PLDMBE}. Obviously, this mechanism works in
any dimension. 

\section{Conclusion}

In the present paper we have studied a model for pulsed laser
deposition with and without Ehrlich-Schwoebel barriers. The model assumes that
the particles are deposited at thermal energies where the transient
mobility of adatoms as well as implantation effects
can be neglected. First we investigated the case
without Ehrlich-Schwoebel barriers. For low pulse intensities PLD
displays essentially the same behavior as MBE. Increasing the
intensity it crosses over to a different behavior 
characterized by the nucleation of many small islands after the first
pulse. The behavior in this regime can be analyzed by studying
the time-dependent nucleation density. Extending previous results we
have shown that in the limit $D/F \to \infty$ this quantity displays an 
unusual type of logarithmic scaling behavior. 

Turning to PLD with Ehrlich-Schwoebel barriers without leakages
we first showed that
in the limit $D/F \to \infty$ the second layer nucleation probability
grows with increasing intensity, enhancing the Villain
instability. This means that in this limit PLD cannot improve
layer-by-layer growth. For finite $D/F$, however, the situation is
different. Studying the surface width for the Schwoebel lengths
between $1$ and $10^4$ we found that PLD produces a smoother surface
than MBE if the barrier is strong enough. This reversal can be explained
by the influence of the initial nucleation density after the first
pulse on the roughening due to the Villain instability.
However, we believe that 
this mechanism alone cannot explain the experimentally observed 
improvement of layerwise growth in PLD where the barriers are much
lower. In order to describe this effect more accurately the transient
mobility of adatoms would have to be taken into account.

As we pointed out, Schinzer {\it et al.}~\cite{Biehl} had reached a similar
conclusion that pulsed deposition leads to increased
roughness. However, this statement, which contradicts an idea by
Rosenfeld {\it et al.}~\cite{Rosenfeld} cannot be universally valid, as our
more detailed investigation showed. In our opinion the negative effect
of pulses is mainly due to the fact that the probability of multiple
deposition on top of islands is increased. This
indicates, that the very high intensity of the single pulse per
monolayer was responsible for the roughness in the simulations of
Schinzer {\it et al.}~\cite{Biehl}. 

However, Schinzer {\it et al.}~\cite{Biehl} also discovered a way to {\em
improve} MBE with pulsed deposition: If desorption cannot be
neglected, atoms will preferentially evaporate from the top
terraces. Thereby the negative effect of multiple deposition can be
compensated, so that one gains the benefit of enforcing small
islands by the pulse in the beginning of the monolayer.

Our model describes only a limiting case of pulsed laser deposition.
It has the virtue of making new scaling concepts clear. In practice
the atoms are seldomly deposited with thermal energy in pulsed laser
deposition. In particular it would be interesting to work out the 
effect of transient mobility due to heating by the pulse.\\[2mm]

\noindent
{\bf Acknowledgements:}\\
We would like to thank J. Krug for pointing out Refs.~\cite{Kadanoff88,Tang89}. 
We also thank the Deutsche Forschungsgemeinschaft for support within SFB 491.

\appendix\section{Finite-Size effects in the limit of low intensity}
\label{sec:finite_gamma}
%
%
\begin{figure}[t]
\vspace*{-8mm}
\includegraphics[width=70mm, angle=270]{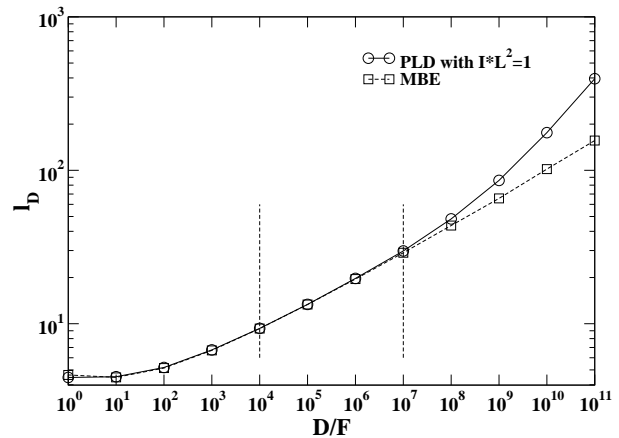}
\caption{Island distance for MBE and PLD with $I\cdot L^2=1$ 
as a function of $D/F$ for a $400\times 400$ lattice. 
The island distance is measured at $0.2$ML.}
\label{FigGamma}
\end{figure}
\begin{figure}
\includegraphics[width=70mm, angle=270]{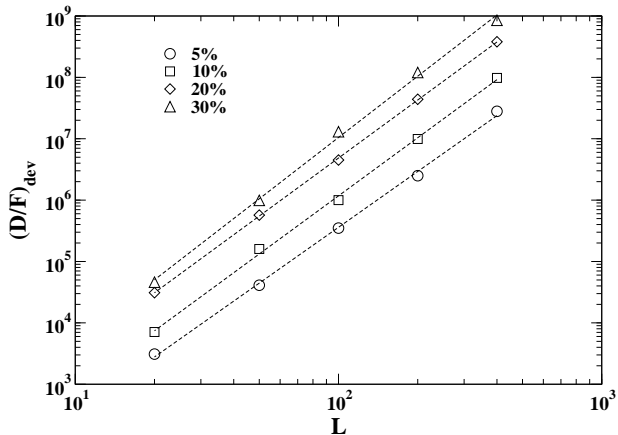}
\caption{Values of $D/F$ where the island distances in PLD 
with $IL^2=1$ and in MBE differ at least by a certain 
percentage specified in the legend. The slopes of the 
power-law fits from bottom to top are estimated by 
$\chi=3.0(2)$, $3.1(2)$, $3.1(2)$, and $3.2(2)$.}
\label{FigFiniteGamma}
\end{figure}
As discussed in Sec.~\ref{SecMBEPLD}, for small intensities
 PLD displays essentially the same growth
behavior as MBE. However, on finite lattices both processes are not
exactly equivalent. This can be seen in Fig.~\ref{FigGamma}, where the
average island distances for MBE and for PLD with an intensity of one
particle per pulse $I=L^{-2}$ are compared over a wide range of $D/F$. For very
low values of $D/F$ the curves coincide but do not follow a power law
because of lattice effects. Only in the range $10^4\leq D/F \leq 10^7$
the curves follow approximately the expected power law $\ell_D\propto
(D/F)^{\gamma}$. Finally, for $D/F\geq
10^7$ the island distances of MBE and PLD differ increasingly from
each other.

In order to demonstrate that this discrepancy between PLD and MBE is a
finite size effect we determine the values of $(D/F)_{\rm dev}$ 
from where on the
island distances in both models differ by a certain factor. As shown
in Fig.~\ref{FigFiniteGamma} these values increase algebraically with
the system size as 
\begin{equation}
  (D/F)_{\rm dev}\propto L^{\chi} \,,
\label{eq:finite_power}
\end{equation}
where the exponent $\chi$ is found to be close to $3$.
This power law  behavior can be explained as follows. 
In both cases single atoms are deposited and the 
average time between two depositions $\tau = 1/L^2$ is the same. 
However, in PLD the deposition takes place in constant 
time intervals, whereas in MBE atoms are randomly 
deposited so that the time intervals between deposition events
obey a Poissonian distribution $P(\tau)=L^2e^{-L^2 \tau}$ with the variance 
$\sigma=\sqrt{\langle \tau^2 \rangle - \langle \tau \rangle^2 } = 1/L^2$.
If such a fluctuation leads to a $\tau$ which is smaller than the
average time interval between two depositions, more nucleations will
be produced. As the formation of nucleations is irreversible
this enhancement is not compensated by fluctuations 
in opposite direction where $\tau$ is large. 
Therefore, the influence of fluctuations increases the number 
of nucleations, leading to a smaller average island distance 
than in PLD. However, this effect can only be seen if the 
fluctuations are strong enough, i.e., they have to be at 
least of the same order as ${\ell_D}^2/(D/F)$, which is the 
average diffusion time (in ML) before an adatom reaches the 
edge of an island or another adatom. Since 
${\ell_D}\sim(D/F)^\gamma$ this argument leads to the scaling relation
\begin{equation}
 (D/F)_{\rm dev}\propto L^{\chi}\,, \qquad \chi
    =\frac{2}{1-2\gamma}.
\end{equation}
For $\gamma =1/6$ the exponent is $\chi =3$ while for  DLA it is given
by $\chi \simeq3.1$. This result is in fair
agreement with the numerically determined exponents in
Fig.~\ref{FigFiniteGamma}.


\end{document}